\documentclass[reprint,aps,prl]{revtex4-1}
\usepackage{graphicx}
\usepackage{dcolumn}
\usepackage{bm}
\usepackage{graphicx}
\usepackage{amsmath}
\usepackage{times,epsfig,wrapfig}
\usepackage{graphicx}
\usepackage{float}
\usepackage{subfig}
\usepackage[gray]{xcolor}
\usepackage{tikz,pgffor}

\usetikzlibrary{shadows}
\usetikzlibrary{    decorations.pathreplacing,    decorations.pathmorphing}

\begin{document}

\title{Sheath instabilities in Hall plasmas devices}
\author{A. I. Smolyakov}
\email{andrei.smolyakov@usask.ca}
\author{W. Frias}\affiliation{ Department of Physics and Engineering Physics, University of Saskatchewan \\
 %EndAName
116 Science Place \\
Saskatoon, SK S7N 5E2,  Canada}
\author{I. D. Kaganovich}
\author{Y. Raitses}
\affiliation{Pinceton Plasma Physics Laboratory, Princeton University \\
 %EndAName
Princeton, NJ 085543 }
\begin{abstract}
New class instabilities is identified in Hall plasmas in configurations with
open magnetic field lines. It is shown that sheath resistivity results in a
robust instability driven by the equilibrium electric field. It is
conjectured that these instabilities play a crucial role in anomalous
transport in Hall plasmas devices.
\end{abstract}
\received[Received]{\today}
\maketitle

%\homepage{http://www.Second.institution.edu/~Charlie.Author.}


%\email{wpf274@mail.usask.ca}

\volumeyear{year} \volumenumber{number} \issuenumber{number} \eid{identifier}
%\date[Date text]{date}

%\revised[Revised text]{\today}

%\accepted[Accepted text]{date}

%\published[Published text]{date}

%\startpage{101}
%\endpage{102}
%\tableofcontents

\emph{1. Introduction.} \ Plasmas in contact with material surface form a
narrow region of noncompensated positive charge, or sheath, which occurs as
a result of fast losses of negative charge due to the high mobility of
electrons. This basic phenomena is of fundamental importance for various
plasma physics applications. The width of the sheath region, typically of
the order of the electron Debye length, $\lambda _{D}=\left( T_{e}/4\pi
e^{2}n_{0}\right) ^{1/2}$, is normally many orders of magnitude smaller than
the characteristic length scale of the bulk plasma region. Despite such
large length scale separation sheath may critically affect the properties of
bulk plasma, e.g. anomalous (turbulent) transport in magnetically confined
plasmas \cite{RogersPRL2010,CohenNF2007}.

While the turbulent transport in plasmas has been studied for many years and
there have been significant advances in understanding of dynamics of drift
wave turbulence in magnetized plasmas, the complete understanding remains
elusive. One of the less studied field is the anomalous mobility of
electrons in Hall plasmas\cite{KeidarIEEE2006}. Hall plasmas, with a
characteristic geometrical length scale between the ion and electron Larmor
radii, $\rho _{e}\ll L\ll \rho _{i},$ and intermediate range of the
frequencies between ion and electron cyclotron frequencies, $\omega _{ci}\ll
\omega \ll \omega _{ce}$, represent a special, but common and important case
of magnetically controlled plasma. Hall plasmas conditions are typical for
many laboratory and space plasmas, e.g. for some regions in Earth ionosphere
and magnetosphere.

Hall plasma conditions are typical for many technological applications.
Common feature of these devices is presence of stationary externally applied
electric field $\mathbf{E}_{0},$ which is perpendicular to the equilibrium
magnetic field $\mathbf{B}_{0},$ and produce stationary $\mathbf{E}%
_{0}\times \mathbf{B}_{0}$ drift velocity. The ions, due their large Larmor
radius, are unmagnetized and accelerated in the $\mathbf{E}_{0}$ direction,
while the electron collisions lead to a finite current along the $\mathbf{E}%
_{0}$. As a result, quasineutral plasma is accelerated \ along the $\
\mathbf{E}_{0}.$ Such plasma accelerators, typically in coaxial geometry
with radial magnetic field $\mathbf{B}_{0}$, axial $\ \mathbf{E}_{0}$, and
azimuthal drift $\mathbf{E}_{0}\times \mathbf{B}_{0}$ $,$ also used as
technological plasma sources, have recently became a subject of growing
interest due their applications in electric space propulsion, so called Hall
plasma thrusters.

Despite many years of successful operation in space \cite{MorozovRPPv21},
the physics of the Hall thrusters is not completely understood. Ongoing
expansion of applications of Hall thrusters for various space missions of
growing complexities and associated new requirements dictate the need for
better understanding of physical processes in these devices\cite%
{BouchoulePPCF2004}. One of the critical problems, is the nature of the
electron current across the magnetic field. Electrons are nearly
collisionless in typical applications, and collisional values of the current
are too low to explain the experimental data. It has long been known that
Hall thruster plasmas are typically in a turbulent state exhibiting a
multitude of instabilities and fluctuations in the wide range of frequencies%
\cite{ChoueiriPoP2001}. Plasma turbulence was suggested to be the reason for
anomalous electron mobility, however the exact nature of responsible
instabilities remains unclear. At the same time, an alternative mechanism of
electron conductivity created by electron collisions with the walls, so
called near-wall conductivity, was proposed\cite{MorozovRPPv21}. However,
the reflection of electrons from the ideal sheath would be reversible, so
the stochastisation mechanism of near-wall conductivity is unclear. In the
original work\cite{MorozovRPPv21}, it was suggested that sheath fluctuations
would create the required stochastic motion of electron, though the reasons
for such fluctuations are unknown. Experimental data and numerical modeling
suggest\cite{BonifaceApPhysL2006} that a combination of both turbulent
mobility and near-wall conductivity is required to satisfactory describe
experimental behavior.

In this paper, we propose a novel instability mechanism based on coupling of
sound waves in bulk plasma with sheath fluctuations. Sheath dynamics,
included as a boundary condition for main plasma, provides the positive
feedback rendering waves unstable. Such mechanism affects both turbulent
mobility of bulk electrons and near-wall conductivity of electrons reflected
by fluctuating sheath. Recent numerical simulations have also suggested the
presence of sheath fluctuations\cite{TaccognaApPhysL2009,Adam}. \ There has
been a number of previous studies which concentrated on instabilities in
crossed \ $\mathbf{E}_{0}\times \mathbf{B}_{0}$ fields Hall plasmas, which
were caused by the equilibrium electron flow in combination with density and
magnetic field gradients\nocite{EsipchukSovPhysTech1976,KapulkinIEEE2008},
resistive effects\nocite{LitvakPoP2001,LitvakPoP2004t}, and electron
cyclotron effects\nocite{DucrocqPoP2006}. These instabilities may also be
affected by sheath impedance effects and eb operative in addition to the
instability \ of basic sound waves whcih is studied in this paper.

It was noted earlier that boundary conditions at the sheath amounts to a
specific dissipation for bulk plasma, the so called sheath resistivity\cite%
{Kadomtsev_sheath}. Sheath resistivity was shown to lead to some
instabilities of magnetically confined plasmas for fusion applications\cite%
{CohenNF2007,BerkPFB1991} but was not studied in Hall plasmas conditions.
Original concept of sheath resistivity was developed for conducting walls.
For such a wall, the electron and ion currents into the wall should not be
balanced locally. Therefore, local perturbations of plasma density and
potential lead to different ion and electron currents, thus producing a
finite current into the sheath. For the dielectric wall, the current on the
material surface must be zero. However, the total sheath impedance  has a
finite and large reactive( capacitive) part, so it is able to maintain a
finite oscillating current on the quasineutral plasma side even when the
current into the wall is zero. Recognition of this fact allows us to
describe the sheath effects as a boundary condition for bulk plasma
oscillations. It turns out that in this formulation, the resistive part of
the total sheath impedance leads to a robust instability in \ $\mathbf{E}%
_{0}\times \mathbf{B}_{0}$ Hall plasmas.

Principally, the stationary $\mathbf{E}_{0}\times \mathbf{B}_{0}$ flow of
magnetized electrons of the velocity is a powerful source of free energy in
Hall plasma devices. The electron perturbations with phase velocities lower
than the equilibrium flow velocity $\mathbf{v}_{0}=c\mathbf{E}_{0}\times
\mathbf{B}_{0}/B_{0}^{2}$ will have negative energy: the total energy of the
perturbed system is lower than the energy of the unperturbed state with
equilibrium flow. The negative energy perturbations are prone to dissipation
instabilities.\ Dissipation takes the energy away from the negative energy
perturbation thus inducing the further growth of its amplitude. We
illustrate this mechanism with a simple example of the ion sound instability
in Hall plasma with equilibrium electron flow in Section II. \ The growth of
the negative energy modes induced by sheath dissipation is the exactly the
instability mechanism for fluctuations in Hall plasma plasmas studied in
Section V. This instability affects both turbulent mobility and near wall
conductivity. \

\emph{2. Destabilized ion sound waves in Hall plasmas. }Ions are
unmagnetized \ and described by the equation
\begin{equation}
-i\omega m_{i}\mathbf{V}_{i}=-e\nabla\tilde{\phi},  \label{i}
\end{equation}
We consider the dynamics along the magnetic field and in perpendicular
direction separately, so that
\begin{equation}
\nabla=\widehat{\mathbf{y}}\frac{\partial}{\partial y}+\widehat{\mathbf{z}}%
\frac{\partial}{\partial z},
\end{equation}
where $z$ is the direction along the magnetic field, $\widehat{\mathbf{z}}$
is a unit vector along the equilibrium magnetic field, and $\widehat{\mathbf{%
y}}$ is the unit vector in the azimuthal direction. Then ion continuity
equation takes the form
\begin{equation}
-i\omega\tilde{n}_{i}-i\frac{en_{0}}{\omega m_{i}}\frac{\partial^{2}}{%
\partial y^{2}}\tilde{\phi}+\frac{1}{e}\frac{\partial}{\partial z}\tilde{J}%
_{\Vert i}=0.  \label{ni}
\end{equation}
The electrons are magnetized and, in the lowest order, the perpendicular
electron velocity is given by the
\begin{equation}
\mathbf{V}_{\bot e}=\mathbf{V}_{\bot0}+\frac{c}{B_{0}}\mathbf{b\times}\nabla%
\tilde{\phi},
\end{equation}
$\mathbf{V}_{\bot0}=c\mathbf{E}_{0}\times\mathbf{b}/B_{0}^{2}=-cE_{0x}/B_{0}%
\widehat{\mathbf{y}}$ is the equilibrium electron drift due to the axial
electric field $E_{0x}$.

The electrons have large mobility along the magnetic field and have
Boltzmann distribution of density
\begin{equation}
\tilde{n}_{e}=\frac{e\tilde{\phi}}{T_{e}}n_{0}.  \label{b}
\end{equation}%
It is valid for $\omega <k_{z}v_{Te}$ and is found from the electron
momentum balance along the magnetic field
\begin{equation}
en\frac{\partial \tilde{\phi}}{\partial z}-T_{e}\frac{\partial \widetilde{n}%
_{e}}{\partial z}=0,  \label{mb}
\end{equation}%
electrons are assumed to be isothermal $T_{e}=const.$ In the local
approximation, Boltzmann electrons (\ref{b}) and perturbed ion density from (%
\ref{ni}) result in the dispersion relation for ion sound waves
\begin{equation}
\omega ^{2}=\left( k_{z}^{2}+k_{y}^{2}\right) c_{s}^{2}  \label{deq}
\end{equation}
Note that the equilibrium electron drift $\mathbf{V}_{\bot 0}$ does not
result in the Doppler frequency shift because ions are not magnetized here.
In this derivation, we do \ not utilize the information about the electron
current along the magnetic field, though obviously such a current is finite.
It can be found from the electron continuity equation
\begin{equation}
-i(\omega -\omega _{0})\tilde{n}_{e}-\frac{1}{e}\frac{\partial }{\partial z}%
\tilde{J}_{\Vert e}=0.  \label{nec}
\end{equation}%
where $\omega _{0}=\mathbf{k\cdot V}_{\bot 0}$. Using quasineutrality, the
ion and electron continuity equations can be cast into the form that
explicitly shows a contribution of the current along the magnetic field
\begin{equation}
-i\omega _{0}\tilde{n}_{e}-i\frac{en_{0}}{\omega m_{i}}\frac{\partial ^{2}}{%
\partial y^{2}}\tilde{\phi}+\frac{1}{e}\frac{\partial }{\partial z}\tilde{J}%
_{\Vert }=0,  \label{cc}
\end{equation}%
where $\tilde{J}_{\Vert }=\tilde{J}_{\Vert i}+\tilde{J}_{\Vert e}$. \
Together with (\ref{i}), (\ref{b}), and (\ref{nec}), this equation, of
course, will result in (\ref{deq}).

The instability mechanism of the discussed modes in essential way relies on
the stationary electron flow due to the equilibrium electric field $\mathbf{E%
}_{0}$ and presence of dissipation. To highlight the physical mechanism of
the instability we consider a simple model of the ion sound waves in Hall
plasma in presence of electron collisions. Taking into account the electron
collisions, the parallel momentum balance becomes
\begin{equation}
k_{z}\left( \widetilde{\phi }-\frac{T_{e}}{e}\widetilde{n}_{e}\right) -\frac{%
m_{e}\nu _{e}}{e}\left( \widetilde{v}_{ze}-\widetilde{v}_{zi}\right) =0.
\end{equation}%
Using this expression and the electron continuity one has for the density
perturbations
\begin{equation}
\frac{\widetilde{n}_{e}}{n_{0}}=\frac{e\tilde{\phi}}{T_{e}}\frac{1}{1-i\nu
_{e}\left( \omega -\omega _{0}\right) /k_{z}^{2}v_{Te}^{2}}.
\end{equation}%
Together with ion equation and quasineutrality it results in the dispersion
relation for unstable ion sound waves%
\begin{equation}
\omega ^{2}=k^{2}c_{s}^{2}-i\frac{k^{2}c_{s}^{2}}{k_{z}^{2}v_{Te}^{2}}\left(
\omega -\omega _{0}\right) .
\end{equation}%
This instability can be interpreted as the negative energy sound wave, $%
\omega <\omega _{0}$, destabilized by dissipation. For weakly collisional
plasmas, when $\lambda _{e}>L\sqrt{m_{e}/m_{i}}$, the bulk dissipation due
to interquartile collisions, becomes less important compared to the sheath
dissipation \cite{Kadomtsev_sheath}, where $\lambda _{e}$ is the electron
mean free path, and $L$ is the characteristic length scale of the plasma
slab (in the direction perpendicular to the sheath). In Hall thruster
plasmas, sheath dissipation induces unstable global and small scale modes,
which are considered next.

\emph{4. Boundary conditions and sheath impedance.} The sheath effects are
considered by including the perturbed current along the magnetic field into
the sheath. For stationary state, we assume the standard sheath model with
Bohm ion current
\begin{equation}
J_{i0}=enc_{s},  \label{1}
\end{equation}%
and the electron current
\begin{equation}
J_{e0}=-\frac{1}{2\sqrt{\pi}}env_{Te}\exp\left( -e\phi_{0}/T_{e}\right) .
\label{2}
\end{equation}
The ion sound velocity is defined as $c_{s}^{2}=T_{e}/m_{i}.$ In stationary
conditions the total current into the sheath is zero: \thinspace$%
J_{0e}+J_{0i}=0.$

Potential $\tilde{\phi}$, density $\tilde{n},$ and temperature fluctuations $%
\tilde{T}_{e}$ at the sheath boundary produce the perturbations of the
electron and ion current into the sheath

\begin{equation}
\tilde{J}_{i}=\frac{\tilde{n}_{i}}{n_{0}}J_{0i}+\frac{1}{2}\frac{\tilde{T}%
_{e}}{T_{e0}}J_{0i},  \label{dj1}
\end{equation}%
\begin{equation}
\tilde{J}_{e}=\frac{\tilde{n}_{e}}{n_{0}}J_{0e}+\frac{1}{2}\frac{\tilde{T}%
_{e}}{T_{e0}}J_{0e}-\left( \frac{e\tilde{\phi}}{T_{e}}-\Lambda\frac{\tilde {T%
}_{e}}{T_{e0}}\right) J_{0e},  \label{dj2}
\end{equation}
where
\begin{equation}
\Lambda=\ln\left( \sqrt{\frac{m_{i}}{2\pi m_{e}}}\right) .
\end{equation}

In this model we do not consider the processes inside the sheath layer and
do not specify how the finite current into the sheath is accommodated. The
actual dynamics of this current will depend on the surface material and
conditions, e.g. for the metal surface this current can be admitted into the
wall, while for the dielectric the finite current into the sheath will lead
to perturbation and rearrangement of the spatial charge distribution in the
sheath layer and on the surface. The characteristic frequency of sheath
oscillations is of the order of the ion plasma frequency $\omega_{pi}$.
Therefore, as long as the frequency of the perturbations is lower than $%
\omega_{pi}$ and the amplitude is small,
\begin{equation}
e\tilde{\phi}/T_{e}<\omega/\omega_{pi},  \label{3}
\end{equation}
the sheath perturbation will retain its standard structure (\ref{1}) and (%
\ref{2}), and the current at the sheath boundary can be determined as a
small perturbation of (\ref{1}) and (\ref{2}). This condition alternatively
can be viewed as the condition that the characteristic charge admitted into
the sheath as a result of fluctuations $\tilde{J}/\omega$ is small compared
to the charge of the stationary sheath $Q,$
\begin{equation}
\tilde{J}/\omega<Q.
\end{equation}
Then, the sheath will only experience a small perturbation and the current
at the sheath boundary can be calculated from (\ref{1}) and (\ref{2}). The
charge is the sheath later $Q\simeq\rho_{e}d,$ where the charge density $%
\rho_{e}$ is estimated as $\rho_{e}\simeq\phi_{0}/\left( 4\pi d^{2}\right) $%
, where the sheath thickness is of the order of the Debye length, $%
d\simeq\lambda_{De}$. Assuming that $\tilde{J}\simeq J_{0}e\tilde{\phi}%
/T_{e} $, one obtains the condition
\begin{equation}
\frac{e\tilde{\phi}}{T_{e}}<\frac{e\phi_{0}}{T_{e}}\frac{\omega\lambda_{De}}{%
c_{s}},
\end{equation}
which is equivalent to (\ref{3}).

\emph{5. Global sheath induced modes.} Presence of the sheath at the plasma
boundary imposed additional constraints on plasma dynamics via the
conditions (\ref{dj1}) and (\ref{dj2}). As a result of these constraints,
the ion sound dynamics is modified and new modes appear. In general, these
modes have the eigen-mode structure that depends both on the $z-$ coordinate
and the perpendicular direction $y$. It is instructive first to consider the
modes which have no structure along the magnetic field. Such modes can be
derived within simple model equations averaged along the magnetic field
lines. Since these modes have no profile along the magnetic field, $\partial
/\partial z=0$, we will call them the global modes.

Consider the magnetic field line tube between \thinspace$z=-L$ and $z=L$ and
introduce the averaged potential and density
\begin{equation}
\overline{n}\equiv\frac{1}{2L}\int_{-L}^{L}\tilde{n}dl,
\end{equation}%
\begin{equation}
\overline{\phi}\equiv\frac{1}{2L}\int_{-L}^{L}\widetilde{\phi}dl.
\end{equation}
The averaged quasineutrality (current closure) equation (\ref{cc}) becomes
\begin{equation}
-i\omega_{0}\overline{n}-i\frac{en_{0}}{\omega m_{i}}\frac{\partial^{2}}{%
\partial y^{2}}\overline{\phi}-\frac{1}{eL}\ J_{0e}\frac{e\overline{\phi}}{%
T_{e}}=0,  \label{5}
\end{equation}
One mode can be obtained from (\ref{5}), by using the Boltzmann relation
averaged along the magnetic field line tube
\begin{equation}
\overline{n}_{e}=\frac{e\overline{\phi}}{T_{e}}n_{0}.  \label{6}
\end{equation}
Equations (\ref{5}) and (\ref{6}) produce the dispersion equation for the
damped mode
\begin{equation}
\omega=\frac{k_{y}^{2}c_{s}^{2}}{\omega_{0}+i\nu_{sh}},
\end{equation}
where $\nu_{sh}\equiv c_{s}/L$ is the effective "collision frequency"
characterizing the sheath resistivity.

However, the conditions (\ref{b}) and (\ref{6}) is not the most general
solution of the electron momentum balance equation (\ref{mb}).\ Indeed,
integrating (\ref{mb}) along the magnetic field lines one obtains

\begin{equation}
\tilde{n}_{e}\left( z,y,t\right) =\frac{en_{0}}{T_{e}}\left( \tilde{\phi }%
\left( z,y,t\right) +\overline{C}\left( y,t\right) \right) .  \label{ne1}
\end{equation}
The integration constant $C\left( y,t\right) $, which may depend explicitly
on $y$ and $t$ is determined by boundary conditions for perturbations at the
sheath. Then the perturbed electron current at the boundary found from (\ref%
{dj2}) is
\begin{equation}
\widetilde{J}_{e}=\frac{e}{T_{e}}\overline{C}J_{0e}.
\end{equation}
The density perturbation induced by this current can be found from the
electron continuity equation (\ref{nec}) giving%
\begin{equation}
-i(\omega-\omega_{0})\overline{n}_{e}+\nu_{sh}\frac{e\overline{C}}{T_{e}}%
n_{0}=0.  \label{nec2}
\end{equation}
It is important to note that in this process, the plasma potential is also
perturbed.\ This perturbation can be found from (\ref{ne1}) and (\ref{nec2}):%
\begin{equation}
\widetilde{\phi}=-\overline{C}\frac{\omega-\omega_{0}+i\nu_{sh}}{\omega
-\omega_{0}}.
\end{equation}
Then, the perturbed plasma density written in terms of the electrostatic
potential is
\begin{equation}
\frac{\overline{n}_{e}}{n_{0}}=i\,\nu_{sh}\frac{e}{T_{e}}\,\frac {\overline{%
\phi}}{\omega-\omega_{0}+i\nu_{sh}}.  \label{n}
\end{equation}
We have changed the notation for $\widetilde{\phi}$ to $\overline{\phi}$ to
emphasize that this perturbation depends only in $y$ and $t$ and is
independent of $z$.

Alternatively, these plasma density perturbations can be found directly from
the averaged equation (\ref{nec}) and using (\ref{dj2}) for the perturbed
electron current

\begin{equation}
-i\left( \omega-\omega_{0}\right) \overline{n}_{e}-\frac{1}{eL}\left( \frac{%
\overline{n}}{n_{0}}-\frac{e\overline{\phi}}{T_{e}}\right) \ J_{0e}=0,
\end{equation}
which results in the same expression (\ref{n}).

The perturbed ion density is found from the averaged ion continuity equation
\begin{equation}
-i\omega \overline{n}_{i}-i\frac{en_{0}}{\omega m_{i}}\frac{\partial ^{2}}{%
\partial y^{2}}\overline{\phi }+\frac{1}{eL}\frac{\overline{n}}{n_{0}}\
J_{0i}=0,
\end{equation}%
giving%
\begin{equation}
\frac{\overline{n}_{i}}{n_{0}}=\frac{e\,k_{\perp }^{2}\tilde{\phi}}{\omega
(\omega +i\nu _{sh})m_{i}}.  \label{ni2}
\end{equation}%
Using quasineutrality, from Eqs. (\ref{n}) and (\ref{ni2}), we obtain the
following dispersion relation:%
\begin{equation}
\omega ^{2}+i\omega \left( \nu _{sh}+\frac{k_{y}^{2}c_{s}^{2}}{\nu _{sh}}%
\right) -k_{y}^{2}c_{s}^{2}\left( 1+i\frac{\omega _{0}}{\nu _{sh}}\right) =0.
\end{equation}
In absence of the equilibrium drift ($\omega _{0}=0)$, Eq. (\ref{deq2})
describe sound waves damped by sheath resistivity
\begin{equation}
\omega =-ik_{y}^{2}c_{s}^{2}/\nu _{sh}.
\end{equation}

For large $\omega _{0}>\nu _{sh}$, it has an unstable root with
\begin{equation}
\omega =\pm \left( i\omega _{0}k_{y}^{2}c_{s}^{2}/\nu _{sh}\right) ^{1/2}
\end{equation}

\emph{6. Small scale modes. }In this section we consider the sheath
instability by taking into account the eigen-mode structure in the direction
parallel to the magnetic field. The ion continuity and motion equations
result in the following local expression for ion density
\begin{equation}
\tilde{n}_{i}=\frac{en_{0}}{\omega ^{2}m_{i}}k_{\bot }^{2}\tilde{\phi}-\frac{%
en_{0}}{\omega ^{2}m_{i}}\frac{\partial ^{2}}{\partial z^{2}}\tilde{\phi}.
\label{ni_local}
\end{equation}%
The parallel component of the electron equation of motion is%
\begin{equation}
0=en_{0}\frac{\partial \tilde{\phi}}{\partial z}-T_{e}\frac{\partial \tilde{n%
}_{e}}{\partial z},
\end{equation}%
which, assuming constant electron temperature gives, and we assume $\tilde{%
\phi}=\tilde{\phi}\left( y,t\right) $
\begin{equation}
\tilde{n}_{e}=\frac{e\tilde{\phi}}{T_{e}}n_{0}.  \label{ne_local}
\end{equation}

Using quasineutrality and Eqs. (\ref{ni_local}) and (\ref{ne_local}), we have

\begin{equation}
\tilde{\phi}=\frac{k_{\bot}^{2}c_{s}^{2}}{\omega^{2}}\tilde{\phi}-\frac {%
c_{s}^{2}}{\omega^{2}}\frac{\partial^{2}}{\partial z^{2}}\tilde{\phi}
\label{em}
\end{equation}
This is an eigen mode equation for ion sound oscillation. Since ions are not
subject of $\mathbf{E\times B},$ there is no Doppler shift in the mode
frequency. The eigen mode equation (\ref{em}) is solved in plane geometry $\
-L\leq z\leq L$ with boundary conditions at $z=\pm L$. The boundary
conditions are defined by the current into the sheath (\ref{dj1}) and (\ref%
{dj2})
\begin{equation}
\tilde{J}_{z}=\tilde{J}_{zi}+\tilde{J}_{ze}=-\frac{e\tilde{\phi}}{T_{e}}%
J_{0e}.  \label{b1}
\end{equation}
The ion and electron current are found from electron and ion equations of
motion and continuity equations
\begin{align}
\tilde{J}_{zi} & =-\frac{ie^{2}n_{0}}{m_{i}\omega}\frac{\partial\tilde{\phi }%
}{\partial z},  \label{b2} \\
-i\left( \omega-\omega_{0}\right) \tilde{n}_{e}-\frac{1}{e}\frac{\partial }{%
\partial z}\tilde{J}_{ze} & =0  \label{b3}
\end{align}
Boundary conditions (\ref{b1}), (\ref{b2}), and (\ref{b3}) at $z=\pm L$ and (%
\ref{em}) make a full eigenmode problem.

General solution for (\ref{em}) has the form

\begin{equation}
\phi=A\exp\left( ik_{z}z\right) +B\exp\left( -ik_{z}z\right) ,
\end{equation}
where $k_{z}$ satisfy the equation
\begin{equation}
k_{z}^{2}=\frac{\omega^{2}}{c_{s}^{2}}-k_{\bot}^{2}.  \label{dq}
\end{equation}
Using (\ref{b2}) and (\ref{b3}) and boundary condition (\ref{b1}) one gets
the dispersion relation
\begin{equation}
\frac{k_{z}c_{s}}{\omega}\left[ 1-\frac{\omega\left( \omega-\omega
_{0}\right) }{k_{z}^{2}c_{s}^{2}}\right] =\left. \frac{A\exp\left(
ik_{z}z\right) +B\exp\left( -ik_{z}z\right) }{A\exp\left( ik_{z}z\right)
-B\exp\left( -ik_{z}z\right) }\right\vert _{z=\pm L}
\end{equation}

This dispersion relation can be cast for even $A=B$ and odd $A=-B$ modes in
the form
\begin{align}
\frac{k_{z}c_{s}}{\omega}\left[ 1-\frac{\omega(\omega-\omega_{0})}{%
k_{z}^{2}c_{s}^{2}}\right] & =i\tan k_{z}L,  \label{s1} \\
\frac{k_{z}c_{s}}{\omega}\left[ 1-\frac{\omega(\omega-\omega_{0})}{%
k_{z}^{2}c_{s}^{2}}\right] & =-i\cot k_{z}L.  \label{s2}
\end{align}
Consider the long wavelength approximation $k_{z}L\ll1.$

\begin{equation}
\frac{k_{z}c_{s}}{\omega}\left[ 1-\frac{\omega\left( \omega-\omega
_{0}\right) }{k_{z}^{2}c_{s}^{2}}\right] =\frac{1}{ik_{z}L}  \label{ev}
\end{equation}
and for odd modes with $A=-B$%
\begin{equation}
\frac{k_{z}c_{s}}{\omega}\left[ 1-\frac{\omega\left( \omega-\omega
_{0}\right) }{k_{z}^{2}c_{s}^{2}}\right] =ik_{z}L  \label{odd}
\end{equation}

Using the relation for $k_{z}$, the dispersion relation for the even mode in
the long wavelength limit becomes
\begin{equation}
\left( \omega _{0}+i\nu _{sh}\right) \omega =k_{\perp }^{2}c_{s}^{2}.  \notag
\end{equation}%
There is no instability for the even mode in the long wavelength limit.

Odd mode (\ref{odd}) gives the equation for the complex wavector $k_{z}$
\begin{equation*}
k_{z}^{2}c_{s}^{2}=\frac{(\omega-\omega_{0})\omega}{1-i\omega/\nu_{sh}},
\end{equation*}
Using it in the dispersion equation (\ref{dq}) we obtain
\begin{equation}
\omega^{3}+\left( i\nu_{sh}\omega_{0}-k_{\perp}^{2}c_{s}^{2}\right)
\omega-i\nu_{sh}k_{\perp}^{2}c_{s}^{2}=0.  \label{omega_odd}
\end{equation}

\begin{figure}[tbp]
\begin{center}
\includegraphics[width=0.40\textwidth]{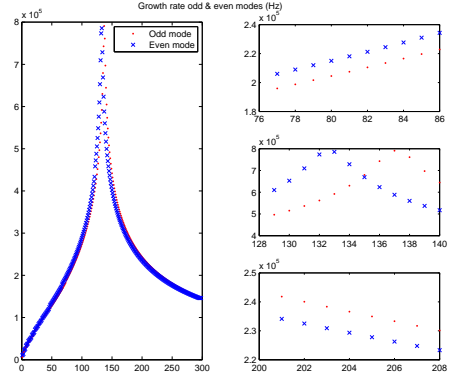}
\end{center}
\caption{Growth rate of the unstable modes in Eqs. () and (). The right
panel shows zoomed in interspacing of even and odd modes.}
\label{}
\end{figure}
\begin{figure}[tbp]
\begin{center}
\includegraphics[width=0.40\textwidth]{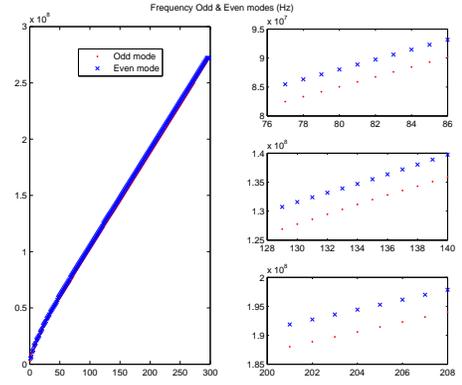}
\end{center}
\caption{Real part of the frequency of unstable modes in Eqs. () and (). The
right panel shows zoomed in interspacing of even and odd modes.}
\label{}
\end{figure}

This has the unstable root instability $\omega =\left( -i\nu _{sh}\omega
_{0}\right) ^{1/2}$ for $k_{\perp }\rightarrow 0\,\ $. For typical Hall
thruster parameters: plasma density $\ n_{0}=10^{12}$ cm$^{-3}$, magnetic
field $B_{0}=150$ G, electric field $E_{0}=$200 V/cm, the electron
temperature $T_{e}=20$ eV, the length scale $L=1$ cm, and lowest wavevector $%
k_{y}$=1.0 cm$^{-1}$, the unstable mode has the frequency and growth rate of
the order of $\omega _{r}\simeq \gamma \simeq $4 MHz. It scales as $\sim
\left( k_{y}E_{0}\right) ^{1/2}$. There is also a weakly unstable root for $%
\omega _{0}\rightarrow 0$ corresponding to a weak evolution of the initial
equilibrium \cite{Kadomtsev_sheath}. The growth rate and real frequency of
the unstable mode in \ref{omega_odd} are shown in Fig. 1.

The small scale instabilities correspond to the roots of the equations (\ref%
{s1}) and (\ref{s2}). The numerical solutions are shown in Fig. 2.

\emph{7. Summary.} To summarize, we have identified novel class of negative
energy wave instability operating in Hall plasma with collisionless
unmagnetized ions and magnetized electrons in presence of the equilibrium
electric field. The instability is triggered by the dissipation due to the
normal current into the sheath. Such instabilities may exist in the form of
the symmetric global mode, Eq. (\ref{omega_odd}), or small scale modes
described by equations (\ref{s1} and (\ref{s2}). The global mode is slowly
varying along the magnetic field, with the characteristic length scale $%
k_{z}1/L$, while small scale modes are characterized by fast variations
along the magnetic field lines. The fluctuations associated with such modes
will affect both turbulent mobility of bulk electrons and near-wall
conductivity of electrons reflected from the fluctuating sheath.

This work is supported in part by NSERC Canada and the Air Force Office of
Scietific\ Research.

\bibliographystyle{unsrt}
\bibliography{sheath}

\end{document}